\documentclass[prd,twocolumn,showpacs,preprintnumbers,aps,nofootinbib]{revtex4-1}

\usepackage{graphicx}
\usepackage{dcolumn}
\usepackage{bm}
\usepackage{amsmath,amssymb,amsfonts}
\usepackage{latexsym}

\usepackage{color}

\def\nn    {\nonumber}

\begin{document}


\title{\boldmath
Implications of  Four-Top and Top-Pair Studies on Triple-Top Production}

\author{Wei-Shu Hou, Masaya Kohda, Tanmoy Modak}
\affiliation{Department of Physics, National Taiwan University, Taipei 10617, Taiwan}
\bigskip


\begin{abstract} 
Multi-top quark production is a staple program at the LHC. 
Single-top and $t\bar t$ productions 
are 
studied extensively, while current efforts are zooming in 
on four-top search, where the Standard Model (SM) 
cross section is at ${\cal O}(10)$ fb. 
In contrast, only at the fb level in SM, triple-top production has not been targeted for study so far.
But such a small cross section makes it a unique probe for New Physics. 
Without the usual discrete $Z_2$ symmetry, 
the general two Higgs doublet model (g2HDM) can naturally
 raise the triple-top production to pb level. 
We illustrate how certain signal regions of four-top search  
can be utilized to constrain triple-top production, 
but urge a dedicated search. 
As an aside, we note that the CMS study at 13 TeV of 
scalar $t\bar t$ resonance interfering with QCD production background
indicate some activity at 400 GeV.
We comment that this could be explained in principle in g2HDM
via the extra top Yukawa coupling. 
\end{abstract}

\maketitle


\section{Introduction}
The top quark was discovered at the Tevatron via 
$t\bar t$ pair production~\cite{Tanabashi:2018oca}, 
and  single-top production was subsequently discovered in 
the $s$-channel~\cite{Giammanco:2017xyn}.
The ATLAS and CMS collaborations quickly rediscovered 
$t\bar t$ production~\cite{Aad:2010ey, Khachatryan:2010ez} 
in early run of the LHC,
and later on discovered single-top in $t$- and $tW$ channels,
and have also studied the $s$-channel~\cite{Giammanco:2017xyn}.
The cross section at 13--14 TeV for 
the QCD initiated $t\bar t$ production is at ${\cal O}(10^3)$~pb, 
while the valence quark initiated $t$-channel single-top production
is more than 200~pb.
In contrast, triple-top production ($tt\bar t$ and $\bar t \bar t t$) 
is predicted\ at the meager few~fb~\cite{Barger:2010uw} level
in the Standard Model (SM), which is negligible compared to $t\bar t$, 
or even single-top. 
As a result, none of the experiments have covered triple-top 
in their search programs so far. 
Although suppressed by four-body phase-space, 
the SM cross section of the QCD initiated four-top ($t\bar t t \bar t$) production, 
at ${\cal O}(10)$~fb, is actually larger than triple-top.
Both ATLAS~\cite{Aaboud:2018jsj} and 
CMS~\cite{Sirunyan:2017roi,Sirunyan:2019wxt} 
have searched for four-top production,
where the more recent CMS study exploits the full Run 2 dataset.

The tiny triple-top cross section makes it 
a good probe for beyond SM (BSM) physics.
It was shown recently~\cite{Kohda:2017fkn}
 (see also Ref.~\cite{Iguro:2017ysu})
that the cross section can reach pb level 
if one drops the usual discrete $Z_2$ symmetry from 
the two Higgs doublet model (2HDM), 
thanks to the presence of extra top Yukawa couplings 
$\rho_{tc}$ and $\rho_{tt}$. 
The additional neutral scalars $H$ or $A$ 
need to be above $t\bar t$ threshold.
Such a general 2HDM (g2HDM) framework allows 
the possibility of approximate alignment without decoupling,
even with $\mathcal{O}(1)$ Higgs quartic couplings~\cite{Hou:2017hiw}.
With $\rho_{tc}$ acting in production and $\rho_{tt}$ in decay, 
the $cg\to t H / t A \to t t \bar t$ processes (conjugate processes always implied)
can enhance triple-top production, making discovery possible 
for semileptonic decays of all three top quarks~\cite{Kohda:2017fkn}.

In this work we illustrate how certain signal regions (SRs) of 
the CMS search for SM production of $t\bar tt\bar t$ at 13 TeV,
based on the datasets of 35.9~fb$^{-1}$~\cite{Sirunyan:2017roi} 
and 137~fb$^{-1}$~\cite{Sirunyan:2019wxt},
can be utilized to constrain triple-top production via $cg\to t H / t A \to t t \bar t$, 
and hence the parameter space for $\rho_{tt}$ and $\rho_{tc}$. 
It is remarkable that SRs not particularly meant for 
triple-top production can give sensitive probes. 
While ATLAS has also searched for four-top production~\cite{Aaboud:2018jsj} 
in the single-lepton and opposite sign dilepton final
states with Run-2 data, we find it less sensitive in our analysis.
Our purpose, however, is to illustrate the need for
a dedicated search for triple-top, 
perhaps as an extension of the four-top search program.

With QCD-induced $t\bar t$ pair production well understood,
it is of interest to consider resonant $t\bar t$ production
interfering with this underlying ``background''.
It has been emphasized~\cite{Carena:2016npr}  
that the rise and dip pattern for (pseudo)scalar channel 
makes the experimental study challenging.
With ATLAS~\cite{Aaboud:2017hnm} leading the way, 
CMS revealed recently their result with Run 2 data~\cite{Sirunyan:2019wph}, 
where there is some hint of activity.
We note that, although it is too early to say, 
the extra top Yukawa coupling $\rho_{tt}$ could in principle be behind this.
The Run 2 study of ATLAS is still missing.
In particular, the full Run 2 data study of both experiments are eagerly awaited.

The paper is organized as follows. 
In Sec.~\ref{form} we discuss briefly the $cg \to tH/tA \to tt\bar t$
triple-top process, followed by CMS four-top search and other
constraints on $\rho_{tc}$ and $\rho_{tt}$ parameter space 
in Sec.~\ref{const}.
We comment on the recent experimental study of 
$gg \to A \to t\bar t$ in Sec.~\ref{disc},
give some further discussion, and end with a summary.

\section{\boldmath 
Triple-Top via $cg\to t H/t A \to t t \bar t$}\label{form}
We consider g2HDM with $CP$-conserving Higgs sector.
The $CP$-even scalars $h$, $H$, the $CP$-odd scalar $A$ and the charged Higgs $H^\pm$
couple to fermions by~\cite{Hou:2017hiw, Davidson:2005cw}
=
\begin{align}
&-\frac{1}{\sqrt{2}} \sum_{F = U, D, L}
 \bar F_{iL} \Big[\big(-\lambda^F_{ij} s_\gamma + \rho^F_{ij} c_\gamma\big) h \nn\\
 &+\big(\lambda^F_{ij} c_\gamma + \rho^F_{ij} s_\gamma\big)H -i ~{\rm sgn}(Q_F) \rho^F_{ij} A\Big]  F_{jR}\nn\\
  &-\bar{U}_i\left[(V\rho^D)_{ij} P_R-(\rho^{U\dagger}V)_{ij} P_L\right]D_j H^+ \nn\\
 &- \bar{\nu}_i\rho^L_{ij} P_R L_j H^+ +{\rm H.c.},
\end{align}
where 
$c_\gamma \equiv \cos\gamma$ and $s_\gamma \equiv \sin\gamma$ 
describe mixing between the two $CP$-even scalars 
($c_\gamma \to 0$ is the alignment limit; see Ref.~\cite{Hou:2017hiw} for definition),
generation indices $i,j =1,2,3$ are summed over, 
$\lambda^F_{ij}=({\sqrt{2}m_i^F}/{v})\, \delta_{ij}$  
and $\rho^F$ are real diagonal and complex $3\times 3$ 
Yukawa coupling matrices, respectively, 
with the vacuum expectation value $v\simeq 246$ GeV.
In particular, we use $\lambda_t = \sqrt{2}m_t/v \simeq 1$ 
with $m_t \simeq 173$ GeV~\cite{Tanabashi:2018oca} throughout the paper.

As discussed earlier, the $cg\to t H/t A \to t t \bar t$ processes are 
induced by $\rho_{tc}$ and $\rho_{tt}$ couplings.
However, non-zero $\rho_{tt}$ and $\rho_{tc}$ induce
 $gg\to H/A\to t \bar t$ and  $gg\to H/A \to t \bar c$, 
as well as $cg\to t H/A \to  t t \bar c$ processes at LHC. 
Although $gg\to H/A\to t \bar t$ is hampered by 
interference with SM $t\bar t$ background~\cite{Carena:2016npr}, 
recent searches by ATLAS~\cite{Aaboud:2017hnm} and CMS~\cite{Sirunyan:2019wph} 
show sensitivity. 
Ref.~\cite{Altunkaynak:2015twa} showed that 
$gg\to H/A \to t \bar c$ should be discoverable at the LHC, 
but the process may suffer from $t+j$ mass resolution, 
which is found close to 200 GeV~\cite{KFC, CMS:2017oas}.
It is not clear whether the latter is  due to the considerably lower cross section 
compared with $s$-channel single-top production.

The $cg\to t H/tA \to  t t \bar c$ process 
with both tops decaying semileptonically gives same-sign top signature 
with low SM background, hence has a unique edge over 
$gg\to H/A\to t \bar t$ and $gg\to H/A\to t \bar c$.
The production of $cg\to t H/tA$ at LHC was first discussed
in Ref.~\cite{Hou:1997pm}, and later in 
Refs.~\cite{Iguro:2017ysu, Kohda:2017fkn, Altmannshofer:2016zrn, Hou:2018zmg}. 
Both $cg\to t H/tA \to t t \bar c$ and $cg\to t H / t A \to t t \bar t$ 
can be discovered at the LHC, where the former may emerge 
perhaps even with Run 2 data~\cite{Hou:2018zmg}. 
Note that both processes can be initiated by $\rho_{ct}$, 
but this coupling is very strongly constrained by
flavor physics~\cite{Altunkaynak:2015twa}.
Non-zero $\rho_{tt}$ may induce the
$gg\to t \bar t H/t\bar tA \to t \bar t t \bar t$ process, 
which is also possible in 2HDM with softly broken $Z_2$ symmetry, 
as discussed in Ref.~\cite{Craig}.
The difference for g2HDM is that $\rho_{tt}$ is complex,
which could drive electroweak baryogenesis~\cite{Fuyuto:2017ewj}.

\section{
Constraints on Triple-Top Production
}\label{const}

In this section we discuss how to constrain the parameter space for
the $\rho_{tc}$ and $\rho_{tt}$ couplings from four-top search. 
But beforehand we first focus on the relevant constraints 
that these two couplings receive individually.
While extracting the constraints from ATLAS and CMS searches, 
we always assume $\rho_{tc}$ and $\rho_{tt}$ are real,
in congruence with the assumptions made by 
the experimental analyses.
However, the impact of complex couplings will be discussed 
later on in the paper.
For simplicity, we set $\rho_{ct} = 0$ and all other $\rho_{ij}=0$, except for $\rho_{tc}$ and $\rho_{tt}$, in our study.

\subsection{General Discussion on Constraints}
\vspace{-0.3cm}
The flavor changing neutral Higgs coupling $\rho_{tc}$ receives 
constraints from both LHC and flavor physics.
In the alignment limit where $c_\gamma = 0$, 
the  strongest constraint on $\rho_{tc}$ arises 
from  the same CMS search for SM four-top production,  Refs.~\cite{Sirunyan:2017roi,Sirunyan:2019wxt}. 
The CRW, i.e. the Control Region for 
$t\bar tW$ background~\cite{Sirunyan:2017roi,Sirunyan:2019wxt} 
provides the most relevant constraint on $\rho_{tc}$. 
For non-zero $\rho_{tc}$, the process $cg\to t H/tA \to t t \bar c$ 
with semileptonically decaying same-sign top contributes abundantly to the CRW region, 
resulting in a stringent constraint on $\rho_{tc}$.
As this has been discussed in Ref.~\cite{Hou:2018zmg},
we refrain from a detailed discussion.
Utilizing the CRW region of Ref.~\cite{Sirunyan:2019wxt} we find
$|\rho_{tc}|\lesssim 0.6$ for  
$m_A = 400$ GeV (or $m_H$), 
while $|\rho_{tc}|\lesssim 0.7$  for 
$m_A = 500$ GeV (or $m_H$) at 
2$\sigma$. 
This should be compared with the limits from the 
CRW of Ref.~\cite{Sirunyan:2017roi}, where the upper limits are
$|\rho_{tc}|\lesssim 0.7$ and $|\rho_{tc}|\lesssim 0.9$ 
for mass of $A$ (or $H$) at 400 and 500 GeV.
Note that the definition of CRW remains unchanged 
between Ref.~\cite{Sirunyan:2017roi} and Ref.~\cite{Sirunyan:2019wxt}, 
while we have assumed $c_\gamma = 0$ and all $\rho_{ij}=0$
except for $\rho_{tc}$.
Due to an exact cancellation between 
the $cg\to t H \to t t \bar c$ and $cg\to t A \to t t \bar c$ 
contributions~\cite{Kohda:2017fkn,Hou:2018zmg}, the constraint 
weakens if $A$ and $H$ 
become degenerate in mass and width.
In such scenarios, $\rho_{tc}$ can still be constrained by
$B_{s,d}$ mixing and $\mathcal{B}(B\to X_s\gamma)$,
where $\rho_{tc}$ enters via charm loop through
$H^+$ coupling~\cite{Crivellin:2013wna} (see also Ref.~\cite{Li:2018aov}).
A reinterpretation of the result from Ref.~\cite{Crivellin:2013wna}
finds $|\rho_{tc}|\lesssim 1.7$ for $m_{H^\pm}=500$ GeV~\cite{ Altunkaynak:2015twa}. Moreover, 
for nonzero $c_\gamma$, the available parameter space 
for $\rho_{tc}$ is strongly constrained by $\mathcal{B}(t\to c h)$. 
The latest ATLAS 95\% CL upper limit based on 36.1 fb$^{-1}$ at 13 TeV 
gives $\mathcal{B}(t\to c h) < 1.1\times 10^{-3}$~\cite{Aaboud:2018oqm}. 
Taking $c_\gamma = 0.2$ for example, one gets
the upper limit of 
{$|\rho_{tc}| \lesssim 0.5$} at $95\%$ CL~\cite{Hou:2019qqi},
but this limit weakens for smaller $c_\gamma$.

The $\rho_{tt}$ coupling can also be constrained by LHC and flavor physics.
For $c_\gamma \neq 0$, $t\bar th$ production 
constrains $\mathcal{O}(1)$ $\rho_{tt}$~\cite{Hou:2018uvr}. 
Regardless of the value of $c_\gamma$, 
$\rho_{tt}$ can be constrained by $B_{s,d}$ mixing, 
but it depends on $m_{H^\pm}$~\cite{Hou:2019grj}. 
For example, $|\rho_{tt}| \gtrsim 1$ is excluded at $95\%$ probability by $B_s$ mixing 
for $m_{H^\pm} = 500$ GeV with $\rho_{ct}=0$~\cite{Altunkaynak:2015twa}.

At this point we note that the ATLAS search for heavy Higgs 
via $gg\to H/A \to t \bar t$~\cite{Aaboud:2017hnm} 
gives more stringent limit on $\rho_{tt}$, even for $c_\gamma=0$. 
The result is based on 20.3 fb$^{-1}$ data at 8 TeV, 
and exclusion limits on $\tan \beta$ vs masses of $H$ and $A$
(for $m_H, m_A > 500$ GeV) are provided in type-II 2HDM framework. 
Assuming $c_\gamma = 0$ and the widths $\Gamma_{H, A}$ are unchanged
from the type-II framework,
we reinterpret the upper limits and find, e.g. $|\rho_{tt}| \lesssim 1$ 
for non-degenerate $H$ or $A$ at 500 GeV, 
while $|\rho_{tt}| \lesssim 0.6$  for $m_A = m_H = 500$ GeV with 95\% CL.
Though keeping silent for a few years,
CMS recently performed a similar search~\cite{Sirunyan:2019wph} 
with 35.9 fb$^{-1}$ data at 13 TeV. Unlike ATLAS which started from 500 GeV, 
CMS searched in the range of
$m_A, m_H=400$--750 GeV, 
and provided model independent 95\% CL upper limit on 
$Att/Htt$ coupling modifiers (see Ref.~\cite{Sirunyan:2019wph} for definition) 
for different values of width vs mass ratios. 
For example, if $\Gamma_A/m_A =5\%$, after reinterpretation of the results, one finds
$|\rho_{tt}|\lesssim 1.1$ ($\lesssim0.9$) for $m_A = 400$ (500) GeV at 95\% CL. 
For a larger $\Gamma_A/m_A =10\%$, the upper limit changes to 
$|\rho_{tt}|\lesssim 1.3$ ($\lesssim 1.0$) for $m_A = 400$ GeV (500 GeV). 
The limits are different for $H$, where $|\rho_{tt}|\lesssim 1.6$ ($\lesssim 1.1$) 
and $|\rho_{tt}|\lesssim 2.1$ ($\lesssim 1.2$)
for $m_H=400$ GeV (500 GeV) for $\Gamma_H/m_H =5\%$ 
and $\Gamma_H/m_H =10\%$  respectively.

However, we note with interest that 
the observed result at $m_A =400$  GeV
has an ``excess'' (wording used by CMS~\cite{Sirunyan:2019wph}) 
with local significance of $\sim 3.5\sigma$ 
when compared with the 95\% CL expected upper limit,
but 1.9$\sigma$ when the look elsewhere effect is taken into account.
We defer a more detailed discussion to Sec.~\ref{disc}.A.

\subsection{Constraint from Four-Top Search}
\vspace{-0.3cm}
We now focus on constraining $\rho_{tc}$ and $\rho_{tt}$, 
and hence on triple-top production, utilizing 
CMS four-top search results,
with data collected 
in 2016~\cite{Sirunyan:2017roi}, 
i.e. 35.9~fb$^{-1}$, and with full Run~2 data~\cite{Sirunyan:2019wxt}, 
i.e. 137~fb$^{-1}$.
Ref.~\cite{Sirunyan:2017roi} divides into 
multiple SRs depending on the number of leptons, 
$b$-tagged jets, with at least two same-sign leptons for the baseline selection criterion.
The search strategy and baseline selection of Ref.~\cite{Sirunyan:2019wxt}, 
where we follow the cut-based analysis, 
are practically the same as Ref.~\cite{Sirunyan:2017roi}, 
with improvements based on taking into account 
full Run~2 detector developments and run conditions.
Ref.~\cite{Sirunyan:2019wxt} improves 
the analysis of Ref.~\cite{Sirunyan:2017roi} 
further by optimizing the definitions of SRs, and adding a few new SRs.
We find SR8 of Ref.~\cite{Sirunyan:2017roi} and SR12 of Ref.~\cite{Sirunyan:2019wxt} as the most relevant 
and provide the most stringent constraints on triple-top production. 
From here on, SR8 will always refer to Ref.~\cite{Sirunyan:2017roi}, 
and SR12 to Ref.~\cite{Sirunyan:2019wxt}.

The selection cuts for SR8 are as follows. 
Each event is required to have at least three leptons ($e,\mu$) 
and at least four jets, with at least three of these $b$-tagged.
The leading lepton transverse momentum ($p_T$) should be $> 25$ GeV, 
while the second lepton with same charge 
and third lepton should have $p_T > 20$ GeV. 
To reduce the background from charge-misidentified Drell-Yan process, 
events with same-sign electron pairs with invariant mass below 12 GeV, 
and events with same-flavor opposite-sign leptons with 
invariant mass below 12 GeV and between 76 GeV and 106 GeV,
 are rejected.
The absolute value of pseudo rapidity ($|\eta|$)
should be $< 2.4$ ($2.5$) for electrons (muons).
The event is selected if $p_T$ of all three $b$-jets are $> 20$ GeV~\cite{info-Jack} 
and the fourth jet with $p_T > 40$ GeV (or 20 GeV if the fourth jet is $b$-tagged).
The scalar sum of $p_T$ of all jets, $H_T$, should be $> 300$ GeV, 
while the missing $p_T$, or $p_T^{\rm miss}$, should be $> 50$ GeV.

With these selection cuts, CMS reported 2 observed events in SR8,
where the expected total number of events 
(SM backgrounds plus $t\bar t t\bar t$) is $2.1 \pm 0.6$.
With semileptonic decay of all three top quarks, 
the $cg\to t H/t A \to t t \bar t$ process contributes to this SR,
which in turn constrains $\rho_{tc}$ and $\rho_{tt}$ 
hence triple-top production.
The selection criteria for SR12 are the same as SR8, except  
restricting the number of jets to four.
With these selection cuts, CMS observed 2 events
in the cut-based analysis, 
with $2.62 \pm 0.54$ events expected.

We remark that supersymmetry search in similar event topologies 
can in principle constrain $\rho_{tc}$ and $\rho_{tt}$.
However, such analyses now typically require $H_T$ and/or missing energy 
that are too large for our purpose.
The selection criteria could be relaxed with $R$-parity violation,
e.g. ATLAS search~\cite{Aaboud:2017dmy} for squark pair production
in $pp \to \tilde d_R \tilde d_R \to \bar t \bar t \bar b \bar b$ or $\bar t \bar t \bar s \bar s$. 
But the selection cuts are still too strong to give meaningful constraint. 
We note further that the ATLAS search for new phenomena 
in events with same-sign leptons and $b$-jets~\cite{Aaboud:2018xpj}
(36.1 fb$^{-1}$ at 13 TeV) has similar SRs, 
but the cuts are again strong and the selection criteria different, 
such that it does not give relevant constraint for our study.

To find the constraint on $\rho_{tc}$ and $\rho_{tt}$, 
we choose two benchmark masses above $2m_t$ threshold for illustration: 
400 and 500~GeV with two different mass hierarchies for $H$ and $A$.
In the first scenario, we assume $m_A$ and $m_H$
are moderately separated with
$m_A < m_{H^\pm} + m_{W^\mp}$ and $m_A < m_H + m_Z$ to
forbid $A \to H^\pm W^\mp$, $Z H$ decays for simplicity.
We denote this scenario as ``$m_A$ alone'' case. 
In the other scenario, we assume $H$ and $A$ are degenerate in mass 
with $m_A < m_{H^\pm} + m_{W^\mp}$ and call it ``mass degenerate'' case. 
For simplicity, we assume $c_\gamma = 0$ and set all $\rho_{ij}^F = 0$, 
except $\rho_{tc}$ and $\rho_{tt}$.
Under these assumptions, $\Gamma_H$ and $\Gamma_A$ are 
nicely approximated as the sum of $H/A \to t \bar t$ 
and $H/A \to t \bar c,\, \bar t c$ partial widths, 
where we neglect tiny  loop-induced $H/A \to \gamma\gamma,\, gg$ rates.

For $m_A$ alone case, we first estimate 
the $cg\to t A \to t t \bar t$ contribution to SR8 and SR12
for $\rho_{tc} = 1$ and $\rho_{tt} = 1$.
We then demand that the sum of the number of events from $cg\to t A \to t t \bar t$ 
and the expected number of events, 
i.e. the total number of events from SM backgrounds 
and $t\bar t t\bar t$ in Refs.~\cite{Sirunyan:2019wxt,Sirunyan:2017roi}, 
agree with the observed number of events within $2\sigma$ uncertainty 
of the expected number.
We then scale by $|\rho_{tc}|^2\times \mathcal{B}(A \to t \bar t)
/\mathcal{B}(A \to t \bar t)\vert_{\rho_{tc} = 1,\rho_{tt} = 1}$
assuming a narrow width for $A$. 
In this regard, we note that for $\rho_{tc} = 1$ and $\rho_{tt} = 1$, 
the total decay width 
for $m_A = 400$ GeV (500 GeV)
is 28.0~GeV  
(44.9~GeV),
which is about $7\%$ ($9\%$) of the mass,
while $\mathcal{B}(A\to t \bar t)$ is $43.4\%$ ($48.3\%$)
 for $m_A = 400$ GeV (500 GeV).

\begin{figure*}[t]
\center
\includegraphics[width=0.36 \textwidth]{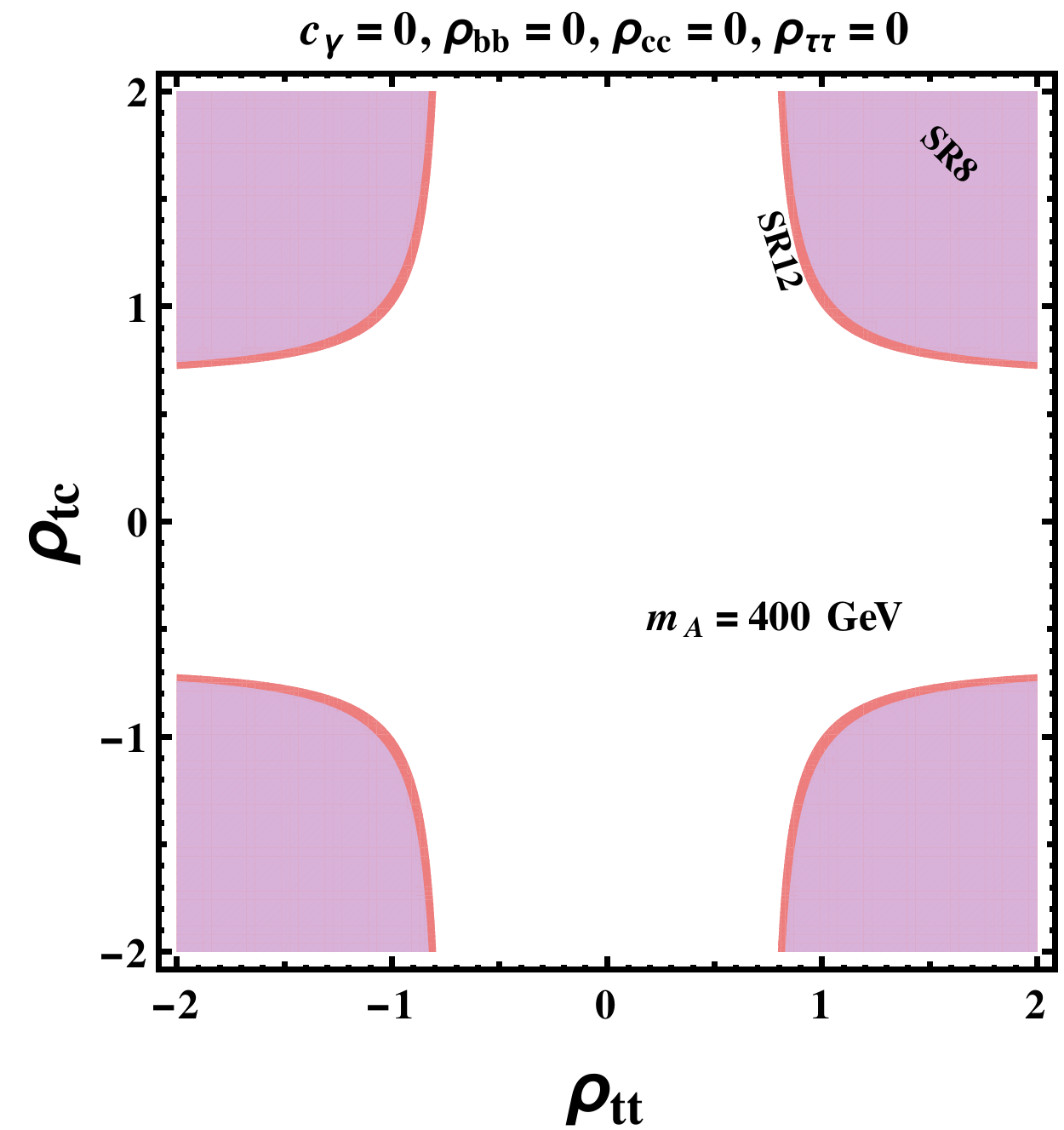}
\includegraphics[width=0.36 \textwidth]{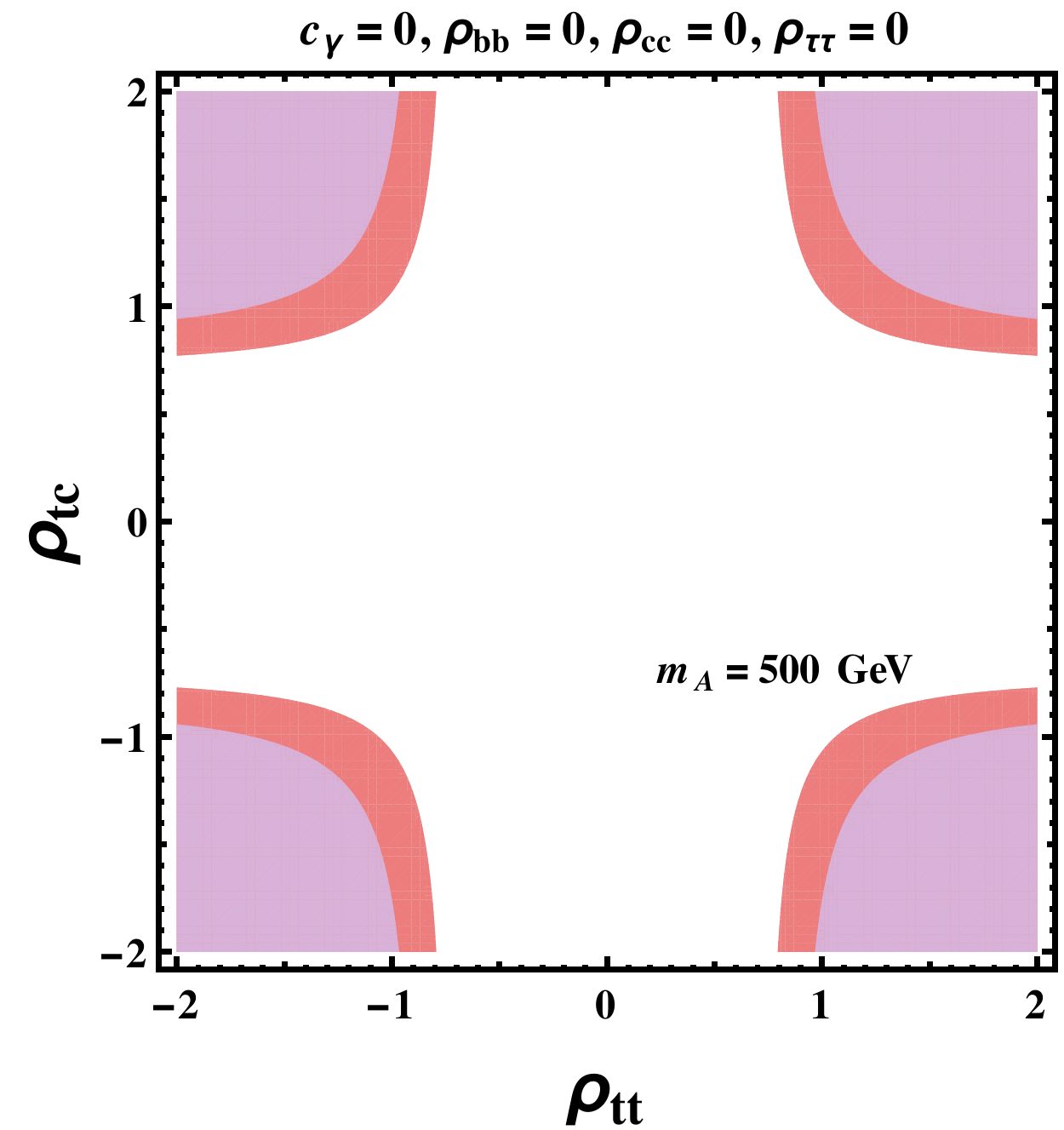}
\caption{
Constraints on $\rho_{tc}$ and $\rho_{tt}$ for $m_A$ alone case, 
extracted from
 SR8 (purple/light shaded) of Ref.~\cite{Sirunyan:2017roi}, and
 SR12 (red/dark shaded) of Ref.~\cite{Sirunyan:2019wxt}, 
for $m_A = 400$ GeV [left] and 500 GeV [right].
}
\label{const_nondeg}
\end{figure*}

\begin{figure*}[t]
\center
\includegraphics[width=0.36 \textwidth]{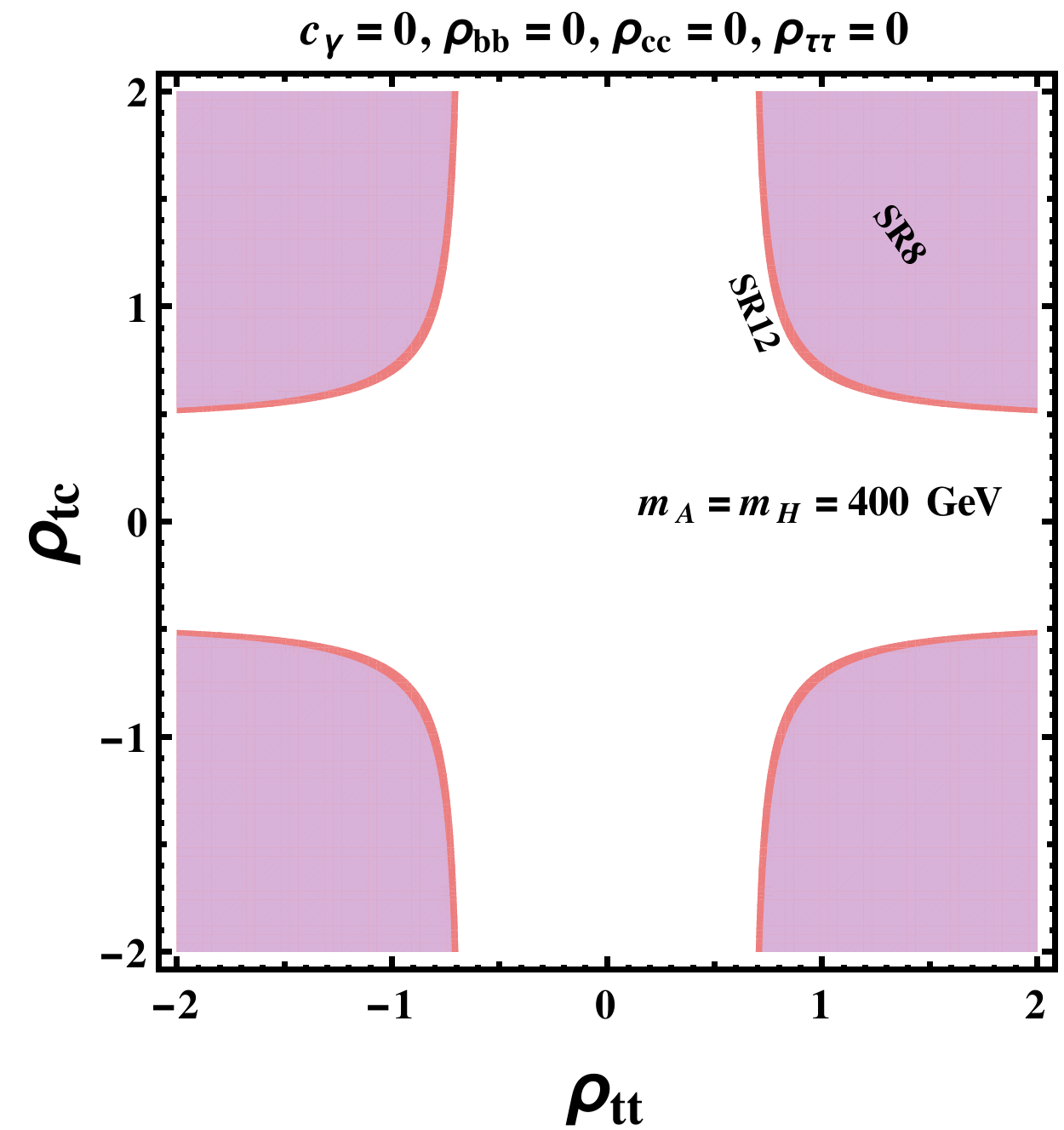}
\includegraphics[width=0.36 \textwidth]{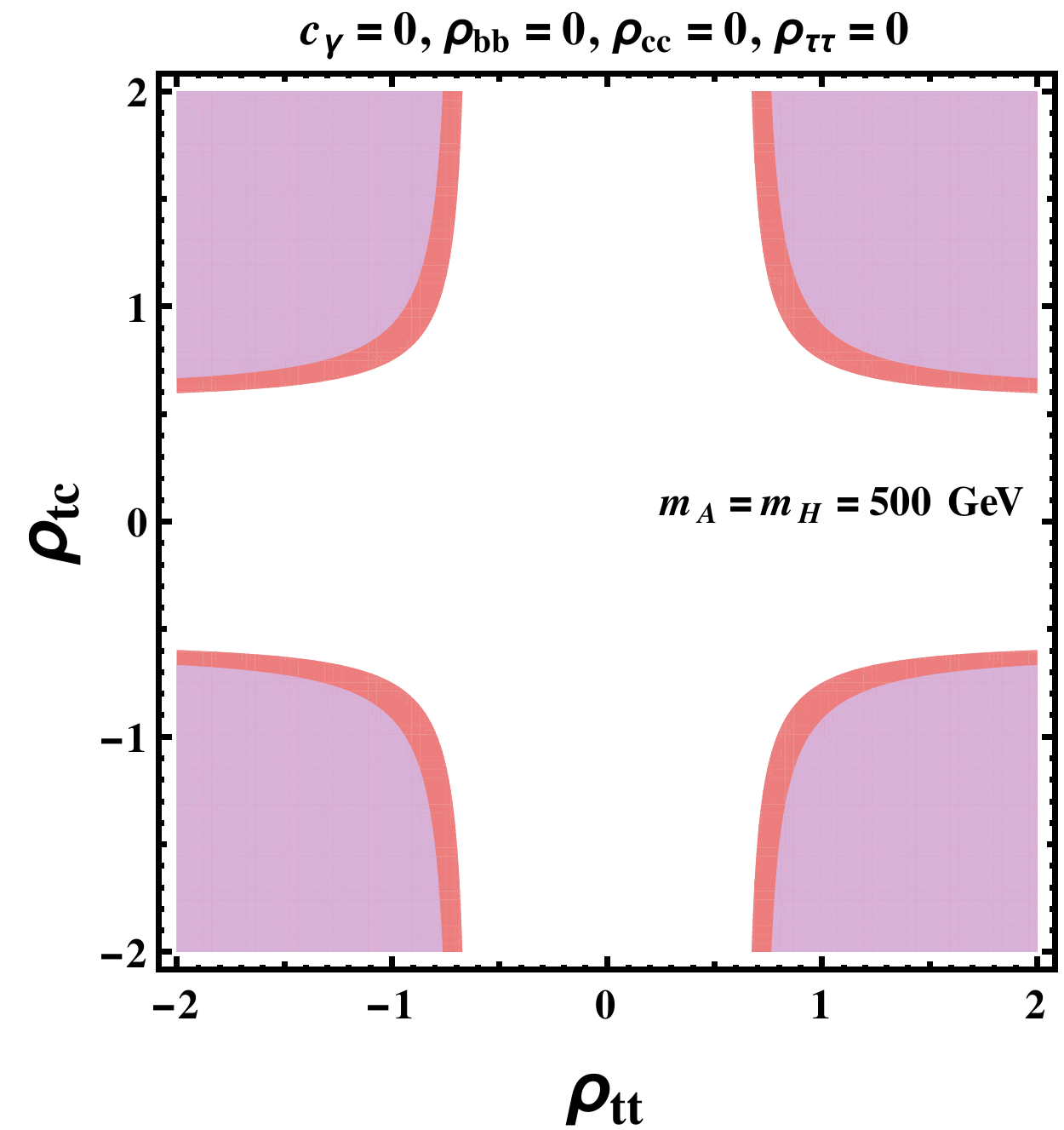}
\caption{
Same as in Fig.~\ref{const_nondeg}, but for mass degenerate case. 
See text for further details.}
\label{const_deg}
\end{figure*}

By simply assuming Gaussian~\cite{excl-poisson} behavior 
for the expected number of events,
the $2\sigma$ exclusion limits are obtained from SR8 and SR12 
for the $m_A$ alone case, as displayed in Fig.~\ref{const_nondeg}
for $m_A = 400$ and $500$ GeV by the purple (light) and red (dark) 
shaded regions, respectively.
We note that, unlike the $cg\to t H \to t t \bar c$ and $cg\to tA \to t t \bar c$ processes, 
the $cg\to t H \to t t \bar t$ and $cg\to t A \to t t \bar t$ processes
do not cancel each other even when $H$ and $A$ are degenerate in mass and width. 
For the mass degenerate case,
the total cross section becomes the incoherent sum of 
$cg\to t H \to t t \bar t$ and $cg\to t A \to t t \bar t$ cross sections. 
A similar scaling strategy is followed for the mass degenerate case, 
and the corresponding $2\sigma$ exclusion limits are shown 
in Fig.~\ref{const_deg} for $m_A = 400$ and $500$~GeV, respectively.

For the $m_H$ alone case, a similar procedure can be followed,
but we note that $\mathcal{B}(H \to t \bar t) = 16.7\%$ 
and $33.0\%$ for $m_H =400$ and 500 GeV respectively, 
which are  smaller than $\mathcal{B}(A \to t \bar t)$ 
for the corresponding masses. 
This is  
because for a real $\rho_{tt}$ with $c_\gamma = \rho_{ct} = 0$, 
$\Gamma (A \to t \bar t) > \Gamma (H \to t \bar t)$, while 
$\Gamma (A \to t \bar c, \bar t c) = \Gamma (H \to t \bar c, \bar t c)$,
for a fixed value of $m_A = m_H$ above the $t\bar c$ threshold.

For numerical results in Figs.~\ref{const_nondeg} and \ref{const_deg},
we generated $pp\to t H/tA + X \to t t \bar t +X$ (with $X$ inclusive) 
at LO for the reference couplings $\rho_{tt}=1$ and $\rho_{tc}=1$, 
utilizing MadGraph5\_aMC@NLO~\cite{Alwall:2014hca} 
with default PDF set NN23LO1~\cite{Ball:2013hta}
for $\sqrt{s}=13$ TeV $pp$ collisions, and interfaced with 
PYTHIA~8.2~\cite{Sjostrand:2014zea} for showering and hadronization. 
We adopt MLM matching~\cite{Alwall:2007fs} prescription
for matrix element and parton shower merging.
The event samples are fed into Delphes~3.4.0~\cite{deFavereau:2013fsa} 
for fast detector simulation following CMS-based detector analysis. 
The effective model is implemented in FeynRules~\cite{Alloul:2013bka}.

Let us elucidate further the results shown in Figs.~\ref{const_nondeg} and \ref{const_deg}.
For $m_A$ alone case, we see from Fig.~\ref{const_nondeg} that 
$|\rho_{tt}| \gtrsim 1\; (1.1)$ is excluded by SR12 for $|\rho_{tc}| \sim 1$ 
and $m_A = 400\, (500)$ GeV, 
The constraint on $\rho_{tt}$ could be relaxed for smaller $|\rho_{tc}|$. 
However, the constraint from $B_s$ mixing or 
ATLAS~\cite{Aaboud:2017hnm} and CMS~\cite{Sirunyan:2019wph} search for heavy Higgs 
via $gg\to H/A\to t \bar t$ could be relevant.
For the mass degenerate case, we see from Fig.~\ref{const_deg} 
that the constraints become stronger,
e.g. $|\rho_{tt}|\gtrsim 0.8$ (0.8) is excluded 
for $m_A = 400$ (500 GeV) for $|\rho_{tc}| > 1$. 
Again, a larger $\rho_{tt}$ is possible, but $\rho_{tc}$ 
would have to be smaller.

It should be stressed that the red (dark) shaded regions 
for SR12 of Ref.~\cite{Sirunyan:2019wxt}  
only mildly surpass the constraints extracted from
purple (light) shaded regions for SR8 of Ref.~\cite{Sirunyan:2017roi}
even though the former is with full Run 2 data.
The primary reason is the more exclusive nature of SR12 compared to SR8, 
i.e. restricting the number of jets to four. 
This highlights the {\it importance of a dedicated search program for 
triple-top production, which is best done by the experiments} themselves, 
and probably just a relatively simple extension 
from the existing four-top search.
It seems to us that the current CMS analysis has reached 
SM sensitivity for four-top,
but would probably need to add Run 3 data to go beyond {\it indication},
i.e. the measured~\cite{Sirunyan:2019wxt} $12.6^{+5.8}_{-5.2}$ fb 
vs the predicted~\cite{Frederix:2017wme} $12.0^{+2.2}_{-2.5}$ fb.
A dedicated triple-top study would be directly probing BSM physics,
as advocated in Ref.~\cite{Kohda:2017fkn}.

\section{Discussion and Summary}\label{disc}

Our original purpose was to urge a dedicated study of triple-top
production by the LHC experiments,
now that four-top search is taking shape.
But in trying to understand the constraint coming from
$gg \to H/A \to t\bar t$, we noticed the mentioning 
of an ``excess'' in the more recent study by CMS, 
Ref.~\cite{Sirunyan:2019wph},
that utilizes Run 2 data taken in 2016.
This ``excess'' is a bit hidden in the CMS Physics Analysis Summary,
i.e. mentioned neither in Abstract, nor in Summary.
We will first  comment on this CMS ``excess'' in the context of g2HDM,
before giving other discussions and our summary.

\subsection{\boldmath Possible ``Excess'' in $gg\to A \to t \bar t$ Search?}\label{excess}
\vspace{-0.3cm}

Searches for $gg \to H/A \to t \bar t$ by ATLAS~\cite{Aaboud:2017hnm} and CMS~\cite{Sirunyan:2019wph} 
are aimed for finding a peak-dip structure in the $t\bar t$ invariant mass ($m_{t\bar t}$) spectrum, 
which distorts from the Breit-Wigner peak by interference with SM $t\bar t$ production.
While ATLAS did not see significant deviation from SM prediction in 8 TeV data~\cite{Aaboud:2017hnm},
CMS reported an $A\to t\bar t$ signal-like deviation frome SM background 
in 35.9 fb$^{-1}$ of  13 TeV data~\cite{Sirunyan:2019wph}.
Based on a model-independent interpretation,
CMS finds the largest deviation for the pseudoscalar $A$ at $m_A =  400$ GeV and $\Gamma_A/m_A = 4\%$ 
with a local significance of $(3.5\pm 0.3) \sigma$ ($1.9\sigma$ with look-elsewhere effect).
It should be noted that the measurement is more involved than 
usual, while the inferred $m_A$ value is rather close to $t\bar t$ threshold.
To conclude the case, one would need a better understanding of 
$t\bar t$ production including the interference with signal near threshold, 
not to mention the need for more statistics.
Nonetheless, it is interesting to contemplate if such a deviation 
can be accommodated within g2HDM.

Although CMS does not provide an estimate for 
the strength of $At\bar t$ coupling (or the coupling modifier $g_{At\bar t}$) 
that corresponds to the $3.5\sigma$ deviation,
we can utilize the material provided~\cite{Sirunyan:2019wph} to infer it.
In particular, the deviation manifests itself in the model-independent constraint plots
as a significant weakening of the observed limit on $g_{At\bar t}$
 at $m_A\sim 400$ GeV from the expected one, 
while no such behavior is seen for scalar $H$.
To illustrate, we take $g_{At\bar t} = 1.1$, 
corresponding to the observed 95\% CL upper limit 
at $m_A = 400$ GeV in the constraint plot for $\Gamma_A/m_A = 5\%$, 
the closest to the reported value of $\Gamma_A/m_A = 4\%$ 
for the 3.5$\sigma$ deviation among the six plots shown,
although the best fit $g_{At\bar t}$ to the deviation
 should be lower~{\cite{info-Jack}}.
We then translate the limit to $\rho_{tt} \simeq 1.1$ 
by the relation $g_{At\bar t} = \rho_{tt}/\lambda_t$.
which we take as a yardstick  in our attempt 
to explain the compatibility of the  deviation with phenomenological constraints.
Note that this is meant only as illustration, 
and we encourage the experiments to provide details of the deviation.

Firstly, we consider the constraint on $\rho_{tt}$ from $pp \to t \bar t A/t \bar t H \to t\bar t t \bar t$ search
by CMS~\cite{Sirunyan:2019wxt},
where 95\% CL upper limits on $\sigma(pp\to (t \bar t , t W , t q) + A/H) \mathcal{B}(A/H\to t \bar t)$
are placed for $m_{A/H} = [350,\, 650]$ GeV.
We utilize the limits to extract 95\% CL upper limits on $\rho_{tt}$, 
assuming $c_\gamma=0$ and all $\rho_{ij}=0$ except $\rho_{tt}$, for simplicity. 
Under these assumptions, the production cross sections of all three processes 
scale as $|\rho_{tt}|^2$ whereas $\mathcal{B}(A\to t \bar t) \simeq 100\%$.
We calculate the three production cross sections by MadGraph5\_aMC@NLO at LO for $\rho_{tt} =1$,
and rescale them.
We then find that $|\rho_{tt}| \gtrsim 0.8$ (0.9) is excluded
for $m_A= 400$ GeV (500 GeV), whereas $|\rho_{tt}| \gtrsim 1.0$ (1.1)
is excluded for $m_H= 400$ GeV (500 GeV).

The apparent tension between the $\rho_{tt} \simeq 1.1$ scenario 
and the CMS $t\bar t A (\to t\bar t)$ search
can be softened in g2HDM with $tcA$ coupling:
 we turn on both $\rho_{tt}$ and $\rho_{tc}$, 
but keep $c_\gamma=0$ for simplicity, so that the limit on $\rho_{tt}$ is alleviated 
by diluting $\mathcal{B}(A \to t \bar t)$ via $A \to t \bar c,\, \bar t c$.
We find that $\rho_{tt}\simeq 1.1$ becomes allowed for $m_A = 400$ GeV
 if $\rho_{tc}\simeq 0.9$.
But $\rho_{tc}$ itself induces $cg \to tA \to tt\bar c$, which is constrained 
by CRW of Ref.~\cite{Sirunyan:2019wxt}, as discussed in the previous section, 
where we found $|\rho_{tc}| \lesssim 0.6$ at 2$\sigma$ 
for $m_A = 400$ GeV with $\rho_{tt} = 0$.
The presence of $\rho_{tt}$ in turn relaxes this constraint by diluting
$\mathcal{B}(A\to t\bar c)$ with $A\to t \bar t$.
We find that $\rho_{tc} \simeq 0.9$ is allowed for $m_A = 400$ GeV if $\rho_{tt}\simeq 1.1$.
The case with $(\rho_{tt},\, \rho_{tc}) \simeq (1.1,\, 0.9)$ and $m_A = 400$ GeV is also allowed 
by the constraints on $cg \to tA \to tt\bar t$ by SR12 and SR8, 
as can be seen in left panel of Fig.~\ref{const_nondeg}.
Intriguingly, the above point is close to the region excluded by SR8 on 
the $(\rho_{tt},\, \rho_{tc})$ plane, 
hence a dedicated triple-top search might be able to probe it.
The $\rho_{tt}$ and $\rho_{tc}$ values translate to 
$\Gamma_A/m_A \simeq 6.9\%$ for $m_A = 400$ GeV.
This is larger than the reported $\Gamma_A/m_A = 4\%$, 
but we note again that the best-fit $\rho_{tt}$ value to the 3.5$\sigma$ deviation 
should be smaller than the value we adopted,
resulting in a smaller $\Gamma_A$.

In the discussion above, 
we have assumed $H$ and $H^{\pm}$ are sufficiently 
heavier than $A$ (with $m_A= 400$ GeV), 
preferably $m_H\gtrsim 500$ GeV and $m_{H^{\pm}}\gtrsim 530$ GeV. 
{We have checked that $(\rho_{tt},\, \rho_{tc}) \simeq (1.1,\, 0.9)$ is allowed 
for $m_H\gtrsim 500$ GeV and satisfies all constraints.}
The choice of $m_{H^{\pm}}\gtrsim 530$ GeV will be discussed in the next paragraph.
To check whether such mass splittings are achievable, we utilize 2HDMC~\cite{Eriksson:2009ws} and 
find that there indeed exists parameter space which 
satisfies perturbativity, tree-level unitarity, and positivity conditions 
as well as oblique $T$ parameter~\cite{Peskin:1991sw} constraint, 
although the Higgs quartic couplings $\eta_i$
 (see Ref.~\cite{Hou:2017hiw} for definition) 
should be sizable, in the range of $3$--$4$. 
Note that if $m_H\sim 500$ GeV, one expects to see 
another peak and dip structure in the $m_{t\bar t}$ mass distribution 
in Refs.~\cite{Aaboud:2017hnm,Sirunyan:2019wxt}.
Such a structure created by $H \to t\bar t$ would merge with 
the tail of the peak-dip structure by $A \to t\bar t$, 
hence, a dedicated analysis simultaneously 
including the $A$ and $H$ effects might be necessary.
Even if the $H$--$A$ mass splitting is large enough to separate the two structures, 
identifying the peak-dip structure by the heavier $H$ would 
face more uncertainties due to limited statistics 
in the falling $m_{t\bar t}$ distributions in higher mass range.

As the $H^+$--$A$ mass splitting cannot be very large, 
one has to take into account the phenomenological constraints 
associated with the $H^+$ boson. 
As discussed in previous section, $\rho_{tt}\simeq 1.1$ 
is allowed by flavor physics 
if $m_{H^\pm} \gtrsim 500$ GeV~\cite{Altunkaynak:2015twa}.
Moreover, we have checked $\rho_{tt}\simeq 1.1$ 
for $m_{H^{\pm}}\gtrsim 530$ GeV is allowed by 
ATLAS search for $gg\to \bar t b H^+ \to \bar t b t \bar b$~\cite{Aaboud:2018cwk},
but similar CMS search~\cite{CMS:1900zym} puts more stringent limit 
and excludes $\rho_{tt}\simeq 1.1$ if all other $\rho_{ij}=0$. 
For this latter search, nonzero $\rho_{tc}$ together with $V_{tb}$ 
gives rise to $\mathcal{B}(H^+ \to c \bar b)$, 
which alleviates the constraint on $\rho_{tt}$. 
We have checked that $\rho_{tt}\simeq 1.1$ is allowed (at 95\% CL) 
for $m_{H^{\pm}} = 530$ GeV if $\rho_{tc}\simeq 0.9$, 
and have considered $H^+ \to A W^+$ decay. 
We thus find that the scenario with $(\rho_{tt},\, \rho_{tc}) \simeq (1.1,\, 0.9)$ and $m_A=400$ GeV
for the 3.5$\sigma$ deviation is viable if $m_H\gtrsim 500$ GeV 
and $m_{H^{\pm}}\gtrsim 530$ GeV.

We have illustrated that a finite parameter space exists in g2HDM where 
the 3.5$\sigma$ deviation~\cite{Sirunyan:2019wph} in $A \to t\bar t$ search could arise.
In setting the exclusion limit, Ref.~\cite{Sirunyan:2019wph}
assumed isolation of $A$ (or $H$) from other states 
such as $H$ ($A$) and ${H^{+}}$.
Such assumptions require Higgs quartic couplings to be sizable. 
It would be interesting to see exclusion plots where 
the masses are relatively close to each other. 
Note also that the similar ATLAS search~\cite{Aaboud:2017hnm} 
for Run 1 data did not cover the mass region below 500 GeV. 
It will be interesting to see an ATLAS study with Run 2 data 
extending down to $m_A\sim 400$ GeV.

Finally, we mention another intriguing aspect of g2HDM.
We have assumed $\rho_{tt}$ (and $\rho_{tc}$) to be real
to conform with experimental analysis.
But complex $\rho_{tt}$ is in fact a robust driver~\cite{Fuyuto:2017ewj} 
for electroweak baryogenesis (EWBG), which is a major attraction
for considering g2HDM that naturally possesses extra Yukawa couplings.
For $CP$-even exotic boson $H$, imaginary $\rho_{tt}$ 
can make it mimic pseudoscalar $A$ in gluon-gluon fusion (ggF) production,
enhancing the cross section~\cite{H-as-A}.
Thus, the ``excess'' at $m_A \sim 400$ GeV may in fact arise from 
a $CP$-even $H$ with extra top Yukawa coupling that is 
close to purely imaginary.
If such is the case, one could maintain custodial SU(2),
i.e. near degeneracy of $A$--$H^+$ that are heavier than $H$,
which would have larger allowed parameter range~\cite{Hou:2017hiw}
than the case of twisted custodial symmetry that we have illustrated above.
One should not only consider two states, i.e. an $H$ at 400 GeV or so
with an $A$ that is heavier with weaker (due to complex $\rho_{tt}$) 
rise-dip interference pattern, but {\it expect $CP$ violation to be 
exhibited in the detailed interference pattern}.
While this illustrates the richness of g2HDM with 
sub-TeV $H$, $A$ and $H^\pm$ bosons, 
it further strengthens our urge for a dedicated study of triple-top, 
to complement the information from the continuation of
the four-top studies.

\subsection{Miscellany}\label{misc}
\vspace{-0.3cm}
We have only discussed triple-top production initiated by $\rho_{tc}$. 
However, $\rho_{tu}$ provides another mechanism for such signature 
via $ug\to tA/tH \to t t \bar t$ production. 
In general, $\rho_{tu}$ should receive even stronger constraint 
than $\rho_{tc}$ from the CRW of Refs.~\cite{Sirunyan:2017roi,Sirunyan:2019wxt}, 
given that the analyses do not distinguish $u$ and $c$ quarks, 
while valence quark initiated $ug\to tA/tH \to t t \bar u$ 
should contribute even more profoundly
than $cg\to tA/tH \to t t \bar c$ in the CRW region~\cite{Hou:2019qqi}. 
However, despite a stronger constraint, 
the discovery potential of $ug \to tA/tH \to t t \bar t$ 
could still be compensated by up-quark PDF enhancement 
compared to $\rho_{tc}$ case.
Note that cancellation between $ug\to t A\to t t \bar u$ and 
$ug\to t H\to t t \bar u$ may relax the  $\rho_{tu}$ constraint 
in the mass degenerate case to some extent~\cite{Hou:2019qqi}, 
resulting in larger triple-top production. 
To put in broader perspective, the current direct search 
constraint~\cite{Aaboud:2018oqm}
on $tuh$ coupling is not so different from $tch$ coupling,
though the null result might reflect approximate alignment (small $\cos\gamma$),
while $\rho_{tu}$ can provide a critical role in enhancing~\cite{Hou:2019uxa} 
$B \to \mu\nu$ in g2HDM, where the effect can be probed at Belle~II.

While setting the upper limits, both ATLAS and CMS
assumed real couplings of $A$, $H$ to top. 
As already mentioned, such assumption has nontrivial impact
while interpreting the experimental limits within g2HDM. 
Here we give a different aspect. Consider $\rho_{tt}$ as purely imaginary, 
then $\Gamma (H \to t \bar t) > \Gamma (A \to t \bar t)$ as well as 
$\sigma(pp\to t\bar t H) > \sigma(pp\to t\bar t A)$, which is complementary to real $\rho_{tt}$.
In such cases the upper limits would be different from the one found in this paper.
Hence, it would be also useful to have the experimental exclusion limits with complex couplings. 
Such complex couplings could be responsible 
for Baryon Asymmetry of the Universe~\cite{Fuyuto:2017ewj}.

Triple-top production may also arise from right handed (RH) $tcZ'$ coupling 
via $cg\to tZ'$ production~\cite{Hou:2017ozb}, followed by $Z'\to t \bar t$ decay
~\footnote{For other triple-top productions see Ref.~\cite{triple-top}}.
In principle, RH and LH $tuZ'$, and LH $tcZ'$ couplings, 
may all produce triple-top signature, but the constraints are
considerably stronger than RH $tcZ'$ coupling.
The phenomenon of $tZ'$ associated production
has been dubbed the potential ``$P'_5$ anomaly for top''~\cite{Hou:2017exe}. 
The case for $Z'\to t \bar t$ decay is a change of model setup,
which will be studied elsewhere.

\subsection{Summary}\label{sum}
\vspace{-0.3cm}
In this paper we advocate a dedicated search for triple-top
production at the LHC, where the cross section in SM is smaller than 
four-top production.
In the general 2HDM without $Z_2$ symmetry, the extra Yukawa couplings
$\rho_{tc}$ and $\rho_{tt}$ give rise to $cg \to tH/A \to tt\bar t$.

A recent CMS study with full Run~2 dataset~\cite{Sirunyan:2019wxt}
found indication for four-top production.
Using four-top search results to constrain the parameter space,
we show that the latest analysis is less restrictive than 
an earlier one~\cite{Sirunyan:2017roi} using smaller dataset,
because event selection became more restrictive, 
which illustrates our point for need of dedicated triple-top analysis.
In understanding constraints, we noticed a search for 
resonant $t\bar t$ production by CMS~\cite{Sirunyan:2019wph} 
reported an ``excess'' for $A \to t\bar t$ at $m_A \sim 400$ GeV. 
We find $\rho_{tt} \sim 1$ could possibly account for the deviation, 
and it could be due to a scalar $H$ boson if $\rho_{tt}$ is purely imaginary. 
While too early to tell, this again highlights the need for
a dedicated triple-top study.

\vskip0.2cm
\noindent{\bf Acknowledgments} \
We thank Yuan Chao and Kai-Feng Chen for discussions.
This research is supported by grants MOST 106-2112-M-002-015-MY3,
107-2811-M-002-039, and 107-2811-M-002-3069.



\begin{thebibliography}{99}




\bibitem{Tanabashi:2018oca} 
  M.~Tanabashi {\it et al.} [Particle Data Group],
  Phys.\ Rev.\ D {\bf 98}, 030001 (2018).

%
\bibitem{Giammanco:2017xyn} 
  A.~Giammanco, R.~Schwienhorst,
  Rev.\ Mod.\ Phys.\  {\bf 90}, 035001 (2018).

%
%
%
%
%
%
%



  
  
  
  
  
  
  
\bibitem{Aad:2010ey} 
  G.~Aad {\it et al.} [ATLAS Collaboration],
  Eur.\ Phys.\ J.\ C {\bf 71}, 1577 (2011).
 
\bibitem{Khachatryan:2010ez} 
  V.~Khachatryan {\it et al.} [CMS Collaboration],
  Phys.\ Lett.\ B {\bf 695}, 424 (2011).
  
  
      
%
\bibitem{Barger:2010uw}
  V.~Barger, W.-Y.~Keung, B.~Yencho,
  Phys.\ Lett.\ B {\bf 687}, 70 (2010).
  


\bibitem{Aaboud:2018jsj} 
  M.~Aaboud {\it et al.} [ATLAS Collaboration],
  Phys.\ Rev.\ D {\bf 99}, 052009 (2019).

\bibitem{Sirunyan:2017roi} 
  A.M.~Sirunyan {\it et al.} [CMS Collaboration],
  Eur.\ Phys.\ J.\ C {\bf 78}, 140 (2018).
  %
%
\bibitem{Sirunyan:2019wxt} 
  A.M.~Sirunyan {\it et al.} [CMS Collaboration],
  arXiv:1908.06463 [hep-ex].
 %
%
%
%
\bibitem{Kohda:2017fkn} 
  M.~Kohda, T.~Modak, W.-S.~Hou,
  Phys.\ Lett.\ B {\bf 776}, 379 (2018).
  
  
\bibitem{Iguro:2017ysu} 
  S.~Iguro, K.~Tobe,
  Nucl.\ Phys.\ B {\bf 925}, 560 (2017).
  
\bibitem{Hou:2017hiw} 
  W.-S.-Hou, M.~Kikuchi,
  EPL {\bf 123}, 11001 (2018).
  
%
  
  
%
\bibitem{Carena:2016npr}
  For a recent reference, see M.~Carena and Z.~Liu,
  JHEP {\bf 1611}, 159 (2016), and references therein.  
%
\bibitem{Aaboud:2017hnm} 
  M.~Aaboud {\it et al.} [ATLAS Collaboration],
  Phys.\ Rev.\ Lett.\  {\bf 119}, 191803 (2017).
%
 %
\bibitem{Sirunyan:2019wph} 
  A.M.~Sirunyan {\it et al.} [CMS Collaboration],
  arXiv:1908.01115 [hep-ex].
%
\bibitem{Davidson:2005cw}
  See, e.g., S.~Davidson and H.E.~Haber,
  Phys.\ Rev.\ D {\bf 72}, 035004 (2005).
  
%
\bibitem{Altunkaynak:2015twa}
  B.~Altunkaynak {\it et al.}, 
  Phys.\ Lett.\ B {\bf 751}, 135 (2015).

\bibitem{KFC}
  K.-F. Chen, private communication.
  See e.g. the $m_{tj}$ resolution in excited top search,
  Ref.~\cite{CMS:2017oas}. 
  
\bibitem{CMS:2017oas} 
  The CMS Collaboration,
  CMS-PAS-B2G-16-025.  
   
\bibitem{Hou:1997pm} 
  W.-S.~Hou, G.-L.~Lin, C.-Y.~Ma, C.-P.~Yuan,
  Phys.\ Lett.\ B {\bf 409}, 344 (1997).

    
    
\bibitem{Altmannshofer:2016zrn} 
  The process was discussed by 
  W.~Altmannshofer {\it et al.},
  Phys.\ Rev.\ D {\bf 94}, 115032 (2016)
  and Ref.~\cite{Iguro:2017ysu},
  but without detailed study.
  See also
  M.~Buschmann, J.~Kopp, J.~Liu, X.-P.~Wang,
  JHEP {\bf 1606}, 149 (2016),
  where the process $ug \to t H^0$ was studied.
%
  The $pp \to t\bar c H^0$ process was discussed by
  S.~Gori, C.~Grojean, A.~Juste , A.~Paul,
  JHEP {\bf 1801}, 108 (2018),
  without detailed study.
%
\bibitem{Hou:2018zmg} 
  An update of Ref.~\cite{Hou:1997pm} was given by
  W.-S.~Hou, M.~Kohda, T.~Modak,
  Phys.\ Lett.\ B {\bf 786}, 212 (2018).

\bibitem{Craig}
  See e.g.
  N.~Craig {\it et al.}, 
  JHEP {\bf 1506}, 137 (2015);
  S.~Kanemura, H.~Yokoya, Y.-J.~Zheng,
  Nucl.\ Phys.\ B {\bf 898}, 286 (2015);
 S.~Gori {\it et al.}, 
  Phys.\ Rev.\ D {\bf 93}, 075038 (2016);
  N.~Craig {\it et al.}, 
  JHEP {\bf 1701}, 018 (2017).
These studies are for 2HDM with softly-broken $Z_2$ symmetry.
%
\bibitem{Fuyuto:2017ewj} 
  K.~Fuyuto, W.-S.~Hou, E.~Senaha,
  Phys.\ Lett.\ B {\bf 776}, 402 (2018).
%
\bibitem{Crivellin:2013wna} 
  A.~Crivellin, A.~Kokulu, C.~Greub,
  Phys.\ Rev.\ D {\bf 87}, 094031 (2013).
 
\bibitem{Li:2018aov} 
  S.-P.~Li, X.-Q.~Li, Y.-D.~Yang,
  Phys.\ Rev.\ D {\bf 99}, 035010 (2019).
%
\bibitem{Aaboud:2018oqm} 
  M.~Aaboud {\it et al.} [ATLAS Collaboration],
  JHEP {\bf 1905}, 123 (2019).
  
  
\bibitem{Hou:2019qqi} 
  W.-S.~Hou, M.~Kohda, T.~Modak,
  Phys.\ Rev.\ D {\bf 99}, 055046 (2019).

  

\bibitem{Hou:2018uvr} 
  W.-S.~Hou, M.~Kohda, T.~Modak,
  Phys.\ Rev.\ D {\bf 98}, 075007 (2018).

%
\bibitem{Hou:2019grj} 
  W.-S.~Hou, R.~Jain, C.~Kao, M.~Kohda, B.~McCoy, A.~Soni,
  arXiv:1901.10498 [hep-ph].
  
  


\bibitem{info-Jack}
  We thank K.-F. Chen for clarifying discussions.   
  
%
\bibitem{Aaboud:2017dmy} 
  M.~Aaboud {\it et al.} [ATLAS Collaboration],
  JHEP {\bf 1709}, 084 (2017).
  
\bibitem{Aaboud:2018xpj} 
  M.~Aaboud {\it et al.} [ATLAS Collaboration],
  JHEP {\bf 1812}, 039 (2018).

%
\bibitem{excl-poisson}
  For more precise estimation of exclusion limits 
  using likelihood function with Poisson counting,
  see e.g.
%
  G.~Cowan, K.~Cranmer, E.~Gross, O.~Vitells,
  Eur.\ Phys.\ J.\ C {\bf 71}, 1554 (2011).
%
  We do not take such a method in our illustrative analysis.  
  
%
\bibitem{Alwall:2014hca}
  J.~Alwall {\it et al.},
  JHEP {\bf 1407}, 079 (2014).
%
\bibitem{Ball:2013hta}
  R.D.~Ball {\it et al.} [NNPDF Collaboration],
  Nucl.\ Phys.\ B {\bf 877}, 290 (2013).
%
\bibitem{Sjostrand:2014zea} 
  T.~Sj\"ostrand {\it et al.},
  Comput.\ Phys.\ Commun.\  {\bf 191}, 159 (2015).
%
\bibitem{Alwall:2007fs}
  J.~Alwall {\it et al.},
  Eur.\ Phys.\ J.\ C {\bf 53}, 473 (2008).
%
\bibitem{deFavereau:2013fsa}
  J.~de Favereau {\it et al.} [DELPHES 3 Collaboration],
  JHEP {\bf 1402}, 057 (2014).
%
\bibitem{Frederix:2017wme} 
  R.~Frederix, D.~Pagani and M.~Zaro,
  JHEP {\bf 1802}, 031 (2018).
%
\bibitem{Alloul:2013bka}
  A.~Alloul {\it et al.}, 
  Comput.\ Phys.\ Commun.\  {\bf 185}, 2250 (2014).
  
\bibitem{Eriksson:2009ws} 
  D.~Eriksson, J.~Rathsman and O.~St{\aa}l,
  Comput.\ Phys.\ Commun.\  {\bf 181}, 189 (2010).
 
\bibitem{Peskin:1991sw} 
  M.E.~Peskin and T.~Takeuchi,
  Phys.\ Rev.\ D {\bf 46}, 381 (1992).
  
\bibitem{Aaboud:2018cwk} 
  M.~Aaboud {\it et al.} [ATLAS Collaboration],
  JHEP {\bf 1811}, 085 (2018).
  
\bibitem{CMS:1900zym} 
  CMS Collaboration [CMS Collaboration],
  CMS-PAS-HIG-18-004.
  
\bibitem{H-as-A}
 Some discussion is given e.g. in Ref.~\cite{Hou:2018uvr}
 for the $\cos\gamma$ component of $h(125)$ in its ggF production.

  
  
%
\bibitem{Hou:2019uxa} 
  W.-S.~Hou, M.~Kohda, T.~Modak, G.-G.~Wong,
  arXiv:1903.03016 [hep-ph].

  
%
\bibitem{Hou:2017ozb} 
  W.-S.~Hou, M.~Kohda, T.~Modak,
  Phys.\ Rev.\ D {\bf 96}, 015037 (2017).
%
\bibitem{Hou:2017exe} 
  G.W.-S.~Hou,
  arXiv:1709.02218 [hep-ph],
  talk presented at DPF 2017, Fermilab, USA.
  

\bibitem{triple-top} 
For a non-exhaustive list, see e.g
  M.~Malekhosseini, M.~Ghominejad, H.~Khanpour and M.~Mohammadi Najafabadi,
  Phys.\ Rev.\ D {\bf 98}, no. 9, 095001 (2018);
%
  C.~Han, N.~Liu, L.~Wu and J.M.~Yang,
  Phys.\ Lett.\ B {\bf 714}, 295 (2012).

\end{thebibliography}
\end{document}